\documentclass[twocolumn,aps,superscriptaddress,showpacs,nofootinbib,floatfix]{revtex4}
\usepackage{epsfig,bm,feynmf}
\usepackage{graphics}
\usepackage{amsmath}
\usepackage[normalem]{ulem}  
\usepackage[dvips]{color} 

\renewcommand\sout{\bgroup \color{red} \ULdepth=-.5ex \ULset}

\begin{document}


\title{Partonic mean-field effects on matter and antimatter elliptic flows}


\author{Taesoo Song}\email{songtsoo@yonsei.ac.kr}
\affiliation{Cyclotron Institute and Department of Physics and Astronomy, Texas A$\&$M University, College Station, TX}
\author{Salvatore Plumari}\email{salvatore.plumari@hotmail.it}
\affiliation{Dipartimento di Fisica e Astronomia, Universit di Catania, Via S. Sofia 64, 95125 Catania, Italy}
\affiliation{Laboratori Nazionali del Sud, INFN, Via S. Sofia 62, 95125 Catania, Italy}
\author{Vincenzo Greco}\email{greco@lns.infn.it}
\affiliation{Dipartimento di Fisica e Astronomia, Universit di Catania, Via S. Sofia 64, 95125 Catania, Italy}
\affiliation{Laboratori Nazionali del Sud, INFN, Via S. Sofia 62, 95125 Catania, Italy}
\author{Che Ming Ko}\email{ko@comp.tamu.edu}
\affiliation{Cyclotron Institute and Department of Physics and Astronomy, Texas A$\&$M University, College Station, TX 77843-3366, USA}
\author{Feng Li}\email{fengphysics@gmail.edu}
\affiliation{Cyclotron Institute and Department of Physics and Astronomy, Texas A$\&$M University, College Station, TX}


\begin{abstract}
Using a partonic transport model based on the Nambu-Jona-Lasinio model, we study the effect of scalar and vector mean fields on the elliptic flows of quarks and antiquarks in relativistic heavy ion collisions in Au+Au collisions at $\sqrt{s_{\rm NN}}=7.7~$GeV and impact parameter $b=8~{\rm fm}$ that leads to the production of a baryon-rich matter. Although the scalar mean field, which is attractive for both quarks and antiquarks, reduces both their elliptic flows, the vector mean field, which is repulsive for quarks and attractive for antiquarks, leads to a splitting of their elliptic flows, and this effect increases with the strength of the vector coupling in the baryon-rich quark matter. Converting quarks and antiquarks at hadronization to hadrons via the quark coalescence model, we further study the dependence of the transverse momentum integrated relative elliptic flow differences between protons and antiprotons, lambda and anti-lambdas, and positively and negatively charged kaons on the strength of the quark vector coupling. These results are then compared with the experimental data measured by the STAR Collaboration in the Beam Energy Scan program at the Relativistic Heavy Ion Collider.
\end{abstract}


\maketitle

\section{introduction}

In relativistic heavy-ion collisions, a hot and dense matter that consists of deconfined quarks and gluons is produced in the initial stage. Depending on the energy of collisions, this so-called quark-gluon plasma (QGP) can have various temperatures and baryon chemical potentials. As the formed QGP expands and cools, it is converted to the normal hadronic matter (HM). Therefore, heavy ion collisions at relativistic energies provide the possibility to study the phase structure of the strongly interacting matter that is described by the quantum chromodynamics (QCD). For the top energy available at the Relativistic Heavy Ion collider (RHIC) and at the Large Hadron Collider (LHC), the produced QGP are nearly baryon free and thus has very small baryon chemical potentials. According to the lattice QCD calculations~\cite{Bernard:2004je,Aoki:2006we,Bazavov:2011nk}, the transition from the QGP phase to the HM phase in this region of the phase diagram is a smooth crossover without a clear phase boundary. The phase transition between QGP and HM is, however, expected to change from the crossover to a first-order transition at certain finite baryon chemical potential called the critical point in the QCD phase diagram~\cite{Asakawa:1989bq,Fukushima:2008wg,Carignano:2010ac,Bratovic:2012qs}. To probe this region of the QCD phase diagram, the Beam Energy Scan (BES) program at much lower energies of $\sqrt{s_{\rm NN}}=7.7$, 11.5, and 39 GeV than the top energy have recently been carried out at RHIC by the STAR Collaboration~\cite{Kumar:2011us}. Although no definitive signals for a first-order phase transition and the critical end point have been established, a number of interesting results have been observed~\cite{Mohanty:2011nm}. One of them is the increasing difference between the elliptic flows of particles and antiparticles, thus a breaking of the constituent quark number scaling of elliptic flows, as the collision energy decreases. Such a behavior cannot be described by a simple hydrodynamic or hadronic cascade model even if the quark coalescence is considered during the hadronization of produced QGP~\cite{Greco:2012hh}. Several theoretical attempts have been made to explain this surprising experimental result~\cite{Dunlop:2011cf,Burnier:2011bf,Xu:2012gf,Steinheimer:2012bn}. In particular, the effect of hadronic mean-field potentials on the elliptic flows of particles and antiparticles has been studied in Ref.~\cite{Xu:2012gf} using a multiphase transport (AMPT) model that includes both initial partonic and final hadronic scatterings~\cite{Zhang:1999bd,Lin:2004en}. Because of the different mean-field potentials for particles and antiparticles in the baryon-rich matter formed in these collisions, $p$, $K^+$ and $\pi^-$ were found to have larger elliptic flows than ${\bar p}$, $K^-$ and $\pi^+$, respectively, as observed in experiments. However, these results are not in quantitative agreement with the experimental data. The calculated relative integrated elliptic flow difference between particles and antiparticles, defined by $[v_2({\rm particle})-v_2({\rm antiparticle})]/v_2({\rm particle})$, for Au+Au collisions at $\sqrt{s_{\rm NN}}=7.7~$GeV and impact parameter $b=8$ fm is about a factor of two smaller than the measured values of 63 \% for $p$ and ${\bar p}$ and -3 \% for $\pi^+$ and $\pi^-$ but is about a factor two larger than the measured value of 13 \% for $K^+$ and $K^-$.
The study in Ref.~\cite{Xu:2012gf} has neglected the effect of mean-field potentials in the initial partonic stage. As shown in Ref.~\cite{Plumari:2010ah} using a partonic transport model based on the Nambu-Jona-Lasinio (NJL) model~\cite{Nambu:1961tp,Nambu:1961fr} for Au+Au collisions at $\sqrt{s_{\rm NN}}=200~{\rm GeV}$, where the baryon chemical potential of produced matter is small, the attractive scalar mean field on quarks and antiquarks was found to reduce their elliptic flows. Since the NJL model is constructed to describe the chiral phase transition, the quark condensate essentially vanishes above the phase transition temperature. The quark scalar mean field is thus small in the quark phase and affects the quark elliptic flow only when the temperature of the system drops below the critical temperature. In the present study, we extend the study of Ref.~\cite{Plumari:2010ah} to also include the effect due to the vector mean field in the quark matter. Since the latter is repulsive for quarks and attractive for antiquarks in baryon-rich quark matter, it increases the elliptic flow of quarks and reduces that of antiquarks. Using the coalescence model to convert quarks and antiquarks to hadrons at hadronization, we further find that the vector mean field in the QGP also leads to an increase of the elliptic flows of $p$ and $\Lambda$, and a decrease of the elliptic flows of $\bar p$ and $\bar\Lambda$. Our results thus demonstrate that besides the hadronic mean-field effects discussed in Ref.~\cite{Xu:2012gf}, including the partonic mean fields also help explain the large $p$ and $\bar p$ as well as $\Lambda$ and $\bar\Lambda$ relative $v_2$ differences observed in experiments. Also, information on the strength of the quark vector mean field is important for understanding the equation of state of QGP at finite baryon chemical potential, which at present cannot reliably be obtained from lattice studies. As shown in studies based on both the NJL and the Polyakov-Nambu-Jona-Lasinio (PNJL) models~\cite{Asakawa:1989bq,Fukushima:2008wg,Carignano:2010ac,Bratovic:2012qs}, the existence and the location of the critical point in the QCD phase diagram is sensitive to the strength of isoscalar vector coupling. A critical point only exists if the strength of the vector coupling is small. Otherwise, the phase transition is always a crossover.

It is worthwhile to note that the elliptic flow in heavy ion collisions at a similar energy range was previously studied in fixed target experiments at both the Alternating Gradient Synchrotron (AGS) and the Super Proton Synchrotron (SPS).  However, the splitting of the particle and antiparticle elliptic flows were not specifically addressed in these studies. Theoretically, a number of transport models were used to study heavy ion collision at these energies, such as the relativistic quantum molecular dynamics (RQMD)~\cite{Sorge:1989vt}, the relativistic transport model (ART)~\cite{Li:1995pra},  the ultra-relativistic quantum molecular dynamics (UrQMD)~\cite{Bass:1998ca}, and the hadron-string dynamics (HSD)~\cite{Cassing:1999es}, but none of them includes a partonic stage in the evolution of the produced hot dense matter. Although the PHSD~\cite{Cassing:2009vt}, which is an extension of the HSD, included the partonic mean fields, it only has the scalar one because it is based on the use of a temperature-dependent parton thermal mass to fit the lattice equation of state at zero baryon chemical potential.

This paper is organized as follows: In Sec.~\ref{NJL}, we briefly review the NJL model for three flavors of quarks and antiquarks and discuss the mean-field approximations. We then describe in Sec.~\ref{RHIC} the partonic transport model, with the mean fields taken from the NJL model, that is used to study the time evolution of the parton phase-space distributions in relativistic heavy-ion collisions. Results on the quark and antiquark elliptic flows from solving the transport model are shown in Sec.~\ref{results} together with those of $p$ and $\bar p$, $\Lambda$ and $\bar\Lambda$, and $K^+$ and $K^-$ that are obtained from quarks and antiquarks by using the coalescence model at hadronization. Finally, a summary is given in Sec.~\ref{summary}.

\section{the Nambu-Jona-Lasinio Model}\label{NJL}

The NJL Lagrangian for three quark flavors has the form~\cite{Bratovic:2012qs}:
\begin{eqnarray}
\mathcal{L}&=&\bar{\psi}(i\not{\partial}-M)\psi+\frac{G}{2}\sum_{a=0}^{8}\bigg[(\bar{\psi}\lambda^a\psi)^2+(\bar{\psi}i\gamma_5\lambda^a\psi)^2\bigg]\nonumber\\
&+&\sum_{a=0}^{8}\bigg[\frac{G_V}{2}(\bar{\psi}\gamma_\mu\lambda^a\psi)^2+\frac{G_A}{2}(\bar{\psi}\gamma_\mu\gamma_5\lambda^a\psi)^2\bigg]\nonumber\\
&-&K\bigg[{\rm det}_f\bigg(\bar{\psi}(1+\gamma_5)\psi\bigg)+{\rm det}_f\bigg(\bar{\psi}(1-\gamma_5)\psi\bigg)\bigg],
\end{eqnarray}
where $\psi=(\psi_u, \psi_d, \psi_s)^T$, $M={\rm diag}(m_u, m_d, m_s)$ and $\lambda^a$ is the Gell-Mann matrices with $\lambda^0$ being identity matrix multiplied by $\sqrt{2/3}$. In the case that the vector and axial-vector interactions are generated by the Fierz transformation of the scalar and pseudo-scalar interactions, their coupling strengths are given by $G_V=G_A=G/2$. The last term is the Kobayashi-Maskawa-t'Hooft (KMT) interaction that breaks $U(1)_A$ symmetry~\cite{'tHooft:1976fv} with ${\rm det}_f$ denoting the determinant in flavor space~\cite{Buballa:2003qv}:
\begin{eqnarray}
{\rm det}_f (\bar{\psi}\Gamma \psi)=\sum_{i,j,k}\varepsilon_{ijk}(\bar{u}\Gamma q_i)(\bar{d}\Gamma q_j)(\bar{s}\Gamma q_k).
\end{eqnarray}
It gives rise to four-point interactions in two flavors and six-point interactions in three flavors. In the two flavor case, the sum of scalar and pseudo-scalar interactions and the KMT interaction with K=-G reduces to the original NJL model~\cite{Asakawa:1989bq,Nambu:1961tp}.

Considering only the flavor singlet vector interaction in the second interaction term and taking the mean-field approximation, the Lagrangian becomes~\cite{Asakawa:1989bq,Hatsuda:1994pi}
\begin{eqnarray}
\mathcal{L}=\bar{\psi}\bigg(i\partial^\mu-\frac{2}{3}G_V\langle\bar{\psi}\gamma^\mu\psi\rangle\bigg)\gamma_\mu\psi-\bar{\psi}M^*\psi+...\,
\label{lagrangian2}
\end{eqnarray}
where $M^*={\rm diag}(M_u, M_d, M_s)$ with
\begin{eqnarray}
M_u&=&m_u-2G\langle\bar{u}u\rangle+2K\langle\bar{d}d\rangle\langle\bar{s}s\rangle,\nonumber\\
M_d&=&m_d-2G\langle\bar{d}d\rangle+2K\langle\bar{s}s\rangle\langle\bar{u}u\rangle,\nonumber\\
M_s&=&m_s-2G\langle\bar{s}s\rangle+2K\langle\bar{u}u\rangle\langle\bar{d}d\rangle,
\label{mass}
\end{eqnarray}
and $...$ denotes constant terms such as $\langle\bar{\psi}\psi\rangle^2$.
The mean fields are calculated as following:
\begin{eqnarray}
\langle\bar{q}_iq_i\rangle=-2M_i N_c\int \frac{d^3{\bf k}}{(2\pi)^3 E_i}~[1-f_i(k)-\bar{f}_i(k)],\nonumber\\
\langle\bar{\psi}\gamma^\mu\psi\rangle=2N_c\sum_{i=u,d,s}\int \frac{d^3{\bf k}}{(2\pi)^3 E_i}~k^\mu[f_i(k)-\bar{f}_i(k)],
\label{condensate}
\end{eqnarray}
where $N_c$ is the number of colors, and $f_i(k)$ and $\bar{f}_i(k)$ are the Fermi-Dirac distributions of partons of flavor $i$ and its anti-flavor, respectively, if the quark matter is in thermal equilibrium. Because the NJL model is not renormalizable, the momentum integration requires a cut-off $\Lambda$. For the parameters in the NJL model, we use those from Refs.~\cite{Bratovic:2012qs,Lutz:1992dv}, that is, $m_u=m_d=3.6$ MeV, $m_s=87$ MeV, $G\Lambda^2=3.6$, $K\Lambda^5=8.9$, and $\Lambda=750$ MeV, unless otherwise stated. We note that the vector coupling in the NJL model can, in principle, be fixed by the masses of $\rho$ and $a_1$ mesons in vacuum. This approach is, however, not reliable as their masses are similar to the cutoff parameter in the NJL model. Also, results from the lattice QCD have led contradictory conclusions on the vector coupling in the partonic matter, with the baryon number susceptibilities requiring a very small vector coupling~\cite{Kunihiro:1991qu,Ferroni:2010xf,Steinheimer:2010sp} and the curvature of the crossover boundary requiring a large one~\cite{Bratovic:2012qs}. Because of these uncertainties, we take the vector coupling as a parameter in the present study.

\section{transport equation for partonic matter}\label{RHIC}

Similar to that for the hadronic matter~\cite{Ko:1987gp,Ko:1988zz}, the time evolution of the partonic matter produced in relativistic heavy ions collisions can be described by the following transport equation for the parton phase-space distribution function $f({\bf x},{\bf p})$;
\begin{eqnarray}\label{transport}
\frac{\partial}{\partial t}f+\vec{v}\cdot\nabla_x f-\nabla_x H\cdot\nabla_p f=\mathcal{C},
\end{eqnarray}
where $H(\bf{x},\bf{p})$ is the Hamiltonian of a quark in the self-consistent scalar and vector mean fields, and $\mathcal{C}$
denotes the collision term that describes the scatterings among partons. Although both the elastic and inelastic scattering
cross sections can be calculated from the NJL model beyond the mean-field approximation~\cite{Hufner:1994vd}, we use in
the present study a constant isotropic parton elastic scattering cross section of 2 mb, which is about a factor of two smaller than the average quark-quark and quark-antiquark elastic cross sections calculated from the NJL model~\cite{Marty:2013ita}, in order to obtain quark and antiquark elliptic flows that, after their hadronization, would lead to hadron elliptic flows similar to those measured in experiments. We note that the use of a smaller parton scattering cross section reflects the fact that treating particles in the corona region of a heavy ion collisions also as partonic matter we overestimates the partonic effect.  Also, the
effect of the annihilation and production of quarks and antiquarks ($q\bar q\leftrightarrow M\bar M$) during the expansion is small in the NJL model since these processes are suppressed at temperature above $T_c$ when meson masses are large compared to those of quarks, which are essentially zero above $T_c$.

For a parton in the scalar and vector mean fields derived from the NJL model, we have in Eq.(\ref{transport})
\begin{eqnarray}
H=\sqrt{M^{*2}+p^{*2}}+g_V\rho^0\equiv E^*\pm g_V\rho^0,
\end{eqnarray}
where ${\bf p}^{*}={\bf p}\mp g_V\mbox{\boldmath$\rho$}$ with $\mbox{\boldmath$\rho$}\equiv\langle\bar{\psi}\mbox{\boldmath$\gamma$}\psi\rangle$ and $g_V \equiv (2/3)G_V$; $\rho^0\equiv\langle\bar{\psi}\gamma^0\psi\rangle$ being the local net baryon density calculated from the parton phase-space distribution function $f(\bf x,p)$. The upper and lower signs are for quarks and antiquarks, respectively.

In the test particle method of solving the transport equation~\cite{Abada:1994mf}, their equations of motion are given by
\begin{eqnarray}
\frac{dx_i}{dt}&=&\frac{\partial H}{\partial p_i}=\frac{p^*_i}{E^*},\label{EoM1}\\
\frac{d p_i}{dt}&=&-\frac{\partial H}{\partial x_i}\nonumber\\
&=&-\frac{M^*}{E^*}\frac{\partial M^*}{\partial x_i} +g_V\bigg(v_j\frac{\partial\rho_j}{\partial x_i}-\frac{\partial\rho_0}{\partial x_i}\bigg).
\label{EoM2}
\end{eqnarray}
In the case that the effective mass of the (anti)quark does not change with position, the LHS of Eq.~(\ref{EoM2}) leads to the familiar Lorentz force:
\begin{eqnarray}
F_i&=&\frac{dp^*_i}{dt}=\frac{dp_i}{dt}-g_V\frac{d\rho_i}{dt}\nonumber\\
&=&g_V\bigg(v_j\frac{\partial \rho_j}{\partial x_i}
-\frac{\partial\rho_0}{\partial x_i}-\frac{\partial\rho_i}{\partial t}-v_j\frac{\partial\rho_i}{\partial x_j}\bigg)\nonumber\\
&=&g_V({\bf v}\times {\bf B}+{\bf E})_i,
\label{Lorentz}
\end{eqnarray}
with ${\bf B}=\nabla\times\mbox{\boldmath$\rho$}$ and ${\bf E}=-\nabla\rho^0-\partial{\mbox{\boldmath$\rho$}}/\partial t$.

Because of the momentum cutoff in Eq.~(\ref{condensate}), only partons with momentum smaller than the cutoff contributes to the scalar and vector densities. In the transport model with densities calculated by dividing the space into cells, this is implemented by counting partons in a cell whose momenta in the rest frame of the cell are less than the cutoff. The mean fields acting on a parton in a cell are then obtained from the scalar and vector densities, the latter being determined from the Lorentz boost of the one calculated in the cell rest frame to the laboratory frame, with corresponding coupling constants if its momentum is below the cutoff. For patrons with momentum above the cutoff, they are not affected by mean fields and are thus treated as free particles. We note that including these high momentum brings the quark matter equation of state closer to that from the lattice QCD calculations, which would otherwise differ significantly~\cite{Marty:2013ita}.

\section{results}\label{results}

\begin{figure}[h]
\centerline{
\includegraphics[width=9 cm]{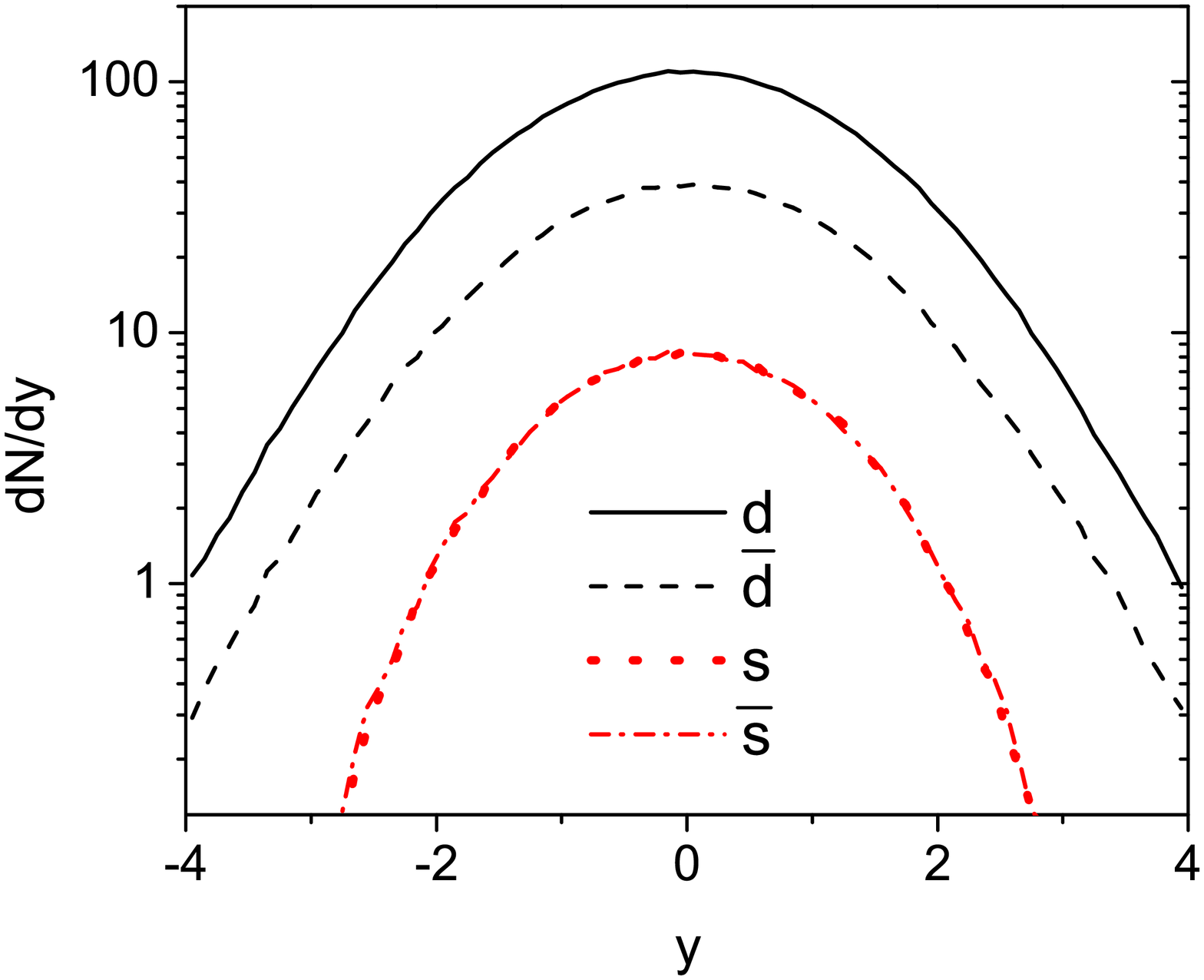}}
\caption{(Color online) Rapidity distributions of down and anti-down quarks as well as (anti-)strange quarks in Au+Au collisions at $\sqrt{s_{\rm NN}}=7.7~$GeV and impact parameter $b=8~{\rm fm}$ from the AMPT model.}
\label{rapidity}
\end{figure}

\begin{figure}[h]
\centerline{
\includegraphics[width=9 cm]{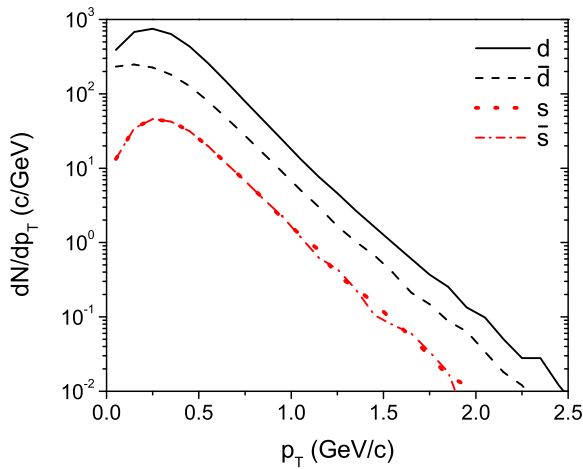}}
\caption{(Color online) Transverse momentum distributions of down and anti-down quarks as well as (anti-)strange quarks in Au+Au collisions at $\sqrt{s_{\rm NN}}=7.7~$GeV and impact parameter $b=8~{\rm fm}$ from the AMPT model.}
\label{momentum}
\end{figure}

For the initial quark and antiquark rapidity and transverse momentum distributions in a relativistic heavy-ion collision, we use the valence quarks and antiquarks converted from hadrons that are obtained from the Heavy-Ion Jet Interaction Generator (HIJING) model~\cite{Gyulassy:1994ew} through Lund string fragmentation as implemented in the AMPT model with string melting~\cite{Lin:2004en}. They are given in Fig.~\ref{rapidity} and Fig.~\ref{momentum} for down and anti-down quarks as well as (anti-)strange quarks in Au+Au collisions at $\sqrt{s_{\rm NN}}=7.7~$GeV and at impact parameter $b=8~{\rm fm}$\footnote{We use this impact parameter since it gives a similar centrality bin as in the experimental analysis~\cite{Adamczyk:2013gw} according to the empirical formula $c=\pi b^2/\sigma_{\rm in}$, where $c$ is the centrality bin and $\sigma_{\rm in} ~ 686$ fm$^2$ is the nucleus-nucleus inelastic cross section from the Glauber model calculation using the nucleon-nucleon inelastic cross sections of about 30.8 mb.}. The distributions of up and anti-up quarks are slightly smaller than those of down and anti-down quarks. It is seen that the number of down quarks is more than twice that of anti-down quarks in heavy ion collisions at this energy, indicating that the produced quark matter is baryon-rich. We note that strange quarks have a softer transverse momentum spectrum than anti-strange quarks in the AMPT model because they are mostly from the conversion of hyperons while anti-strange quarks are mostly from the conversion of strange mesons. The (anti-)strange quark transverse momentum spectrum shown in Fig.~\ref{momentum} is obtained by randomly interchanging the positions and momenta of strange and anti-strange quarks since they are expected to have similar transverse momentum spectra if they were taken directly from the decay of produced strings.

We solve Eqs.~(\ref{condensate}), (\ref{EoM1}), and (\ref{EoM2}) numerically and then use the resulting parton momentum distributions to evaluate their elliptic flow $v_2$ according to
\begin{eqnarray}\label{elliptic}
v_2=\frac{1}{N}\sum_{i=1}^N\frac{(p_x^{*2})_i-(p_y^{*2})_i}{(p_x^{*2})_i+(p_y^{*2})_i},
\end{eqnarray}
where $N$ is the total number of test particles.

Although energy is not conserved in the coalescence model as it is based on the sudden approximation, the violation is, however, not large. As shown in Ref.~\cite{Lin:2004en}, it is about 1\%. With the elliptic flow given by the mean value of the ratio $(p_x^{*2}-p_y^{*2})/(p_x^{*2}+p_y^{*2})$ of produced hadrons as given in Eq.(\ref{elliptic}), results obtained from the coalescence model are expected to be reliable because the momentum is conserved when quarks and antiquarks are combined into hadrons. This can also be inferred from the extended coalescence model introduced in Ref.~\cite{He:2010vw}, which conserves energy by taking into account the widths of hadrons in the hadronic medium, as the quark number scaling of hadron elliptic flow is observed in both the usual coalescence model based on the sudden approximation and the extended coalescence model.

\subsection{quark and antiquark elliptic flows}

\begin{figure}[h]
\centerline{
\includegraphics[width=9 cm]{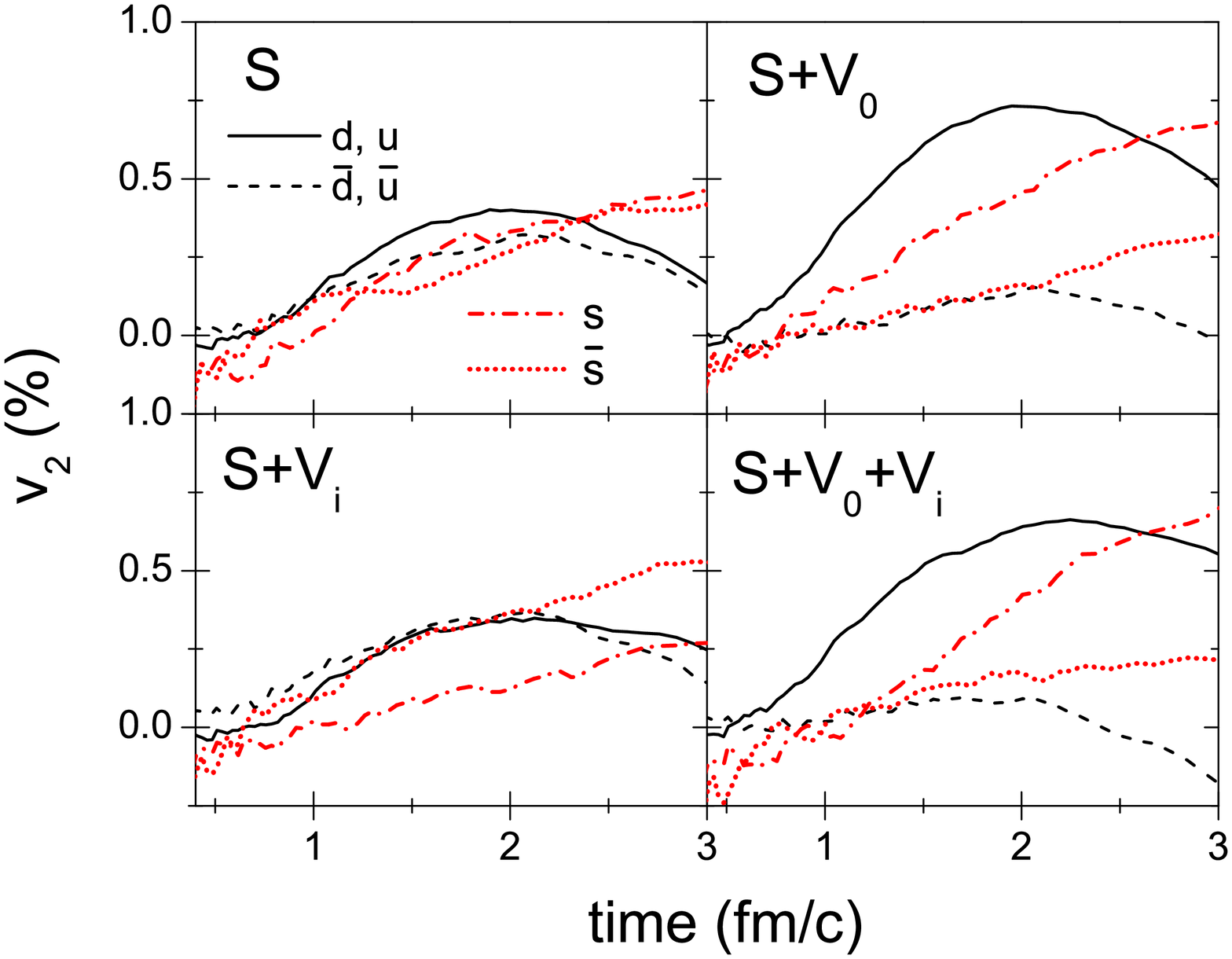}}
\caption{(Color online) Integrated elliptic flow $v_2$ of light and strange quarks and antiquarks at mid-rapidity ($|y|<1$) as functions of time in Au+Au collisions at $\sqrt{s_{\rm NN}}=7.7~$GeV and at impact parameter $b=$ 8 fm for the cases of including only the scalar mean field ($S$), the scalar and the time component of the vector mean field ($S+V_0$), the scalar and the space component of the vector mean field ($S+V_i$), and the scalar and both components of the vector mean field ($S+V_0+V_i$) using $g_V=G/6$.}
\label{v2}
\end{figure}

Figure~\ref{v2} shows the integrated $v_2$ of light and strange quarks and antiquarks as functions of time for the cases of including only the scalar mean field, the scalar and the time component of the vector mean field, the scalar and the space component of the vector mean field, and the scalar and both components of the vector mean field using $g_V=G/6$. It is seen that the difference between quark and antiquark $v_2$ is small if only the scalar mean field is included in determining their propagations in the quark matter. This is due to the fact that through $\nabla M^*$ in Eq.~(\ref{EoM2}) the scalar mean field generates the same attractive force on quarks and antiquarks. Including the time component of the vector mean field caused by excess quarks leads to a larger quark than antiquark $v_2$ due to the resulting repulsive force on quarks and attractive force on antiquarks. We note that $v_2$ is generated by the gradient of the pressure, which always points outward in heavy-ion collisions but is enhanced by a repulsive force and reduced by an attractive force. Therefore, the time-component of the vector mean field enhances the $v_2$ of quarks and suppresses that of antiquarks while the attractive scalar mean field suppresses both quark and antiquark $v_2$~\cite{Plumari:2010ah}. As to the space component of the vector mean field, its effect on the $v_2$ of quarks and antiquarks is opposite to that from the time component but with a much smaller magnitude. The other difference between the effect of the space component and that of the time component of the vector mean field is that the splitting between the $v_2$ of quarks and antiquarks starts earlier in the time-component case than in the space-component case. The reason is that the space component of the vector mean field is proportional to the baryon current that takes time to develop and is not large unless the relativistic flow streams are large~\cite{Greco:2002sp}. The net result of the partonic mean fields in our study is that it leads to a larger quark than antiquark $v_2$ if the produced quark matter is baryon-rich. These behaviors are seen for both light and strange quarks. On the other hand, while the integrated $v_2$ of strange quarks continues to increase with time, that of light quarks decreases at later times. This is due to the stronger attractive scalar mean field for light quarks than for strange quarks as a result of the larger decrease of the light quark condensate along the radial direction than that of the strange condensate and the smaller light quark mass than the strange quark mass.

\begin{figure}[h]
\centerline{
\includegraphics[width=9 cm]{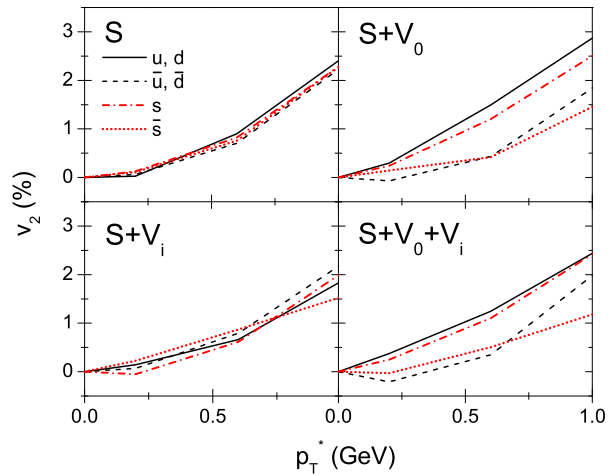}}
\caption{(Color online) Elliptic flow $v_2$ of light and strange quarks and antiquarks at mid-rapidity ($|y|<1$) as functions of transverse momentum at hadronization for the cases of including only the scalar mean field ($S$), the scalar and the time component of the vector mean-field ($S+V_0$), the scalar and the space component of the vector mean field ($S+V_i$), and the scalar and both components of the vector mean field ($S+V_0+V_i$) using $g_V=G/6$.}
\label{v2-pt}
\end{figure}

For the transverse momentum dependence of the quark and antiquark $v_2$, they are shown in Fig.~\ref{v2-pt} for those at the end of the partonic phase, which is about 2.5 fm/c after the start of the partonic evolution when the energy density in the center of produced quark matter decreases to about 0.8 ${\rm GeV/fm^3}$, again for the cases of including only the scalar mean field, the scalar and the time component of the vector mean field, the scalar and the space component of the vector mean field, and the scalar and both components of the vector mean field using $g_V=G/6$. Without the vector mean field, the integrated $v_2$ at the end of the partonic phase is slightly larger for light quarks than for light antiquarks, resulting in a relative $v_2$ difference of about 15~\%, while those of strange and anti-strange quarks are similar as shown in the upper left panel. These results reflect the fact that light quarks, which are mostly from colliding nuclei, has a smaller initial eccentricity than those of produced light antiquarks as well as strange and anti-strange quarks, thus making them more likely to flow in the reaction plane. Including the vector mean field, which has opposite effects on quarks and antiquarks, increases the relative $v_2$ difference between light quarks and antiquarks as well as that between strange and anti-strange quarks as shown in the lower right panel. We note that the vector potential hardly affects the final quark and antiquark transverse momentum spectra in heavy ion collisions at such a high energy as considered here since they are mainly determined by the partonic scattering.


\subsection{hadron elliptic flows}

To study how different quark and antiquark $v_2$ is reflected in the $v_2$ of produced hadrons, we use the coalescence model to convert them to hadrons at hadronization. It is based on the sudden approximation that hadronization occurs fast and hadrons are simply projected out from the parton wave functions. The mean-field potential thus does not play a role in this model, except its effect on the phase space distribution of partons. In this model, the probability for a quark and an antiquark to form a meson is proportional to the quark Wigner function of the meson with the proportional constant given by the statistical factor $g_M$ for colored spin-1/2 quark and antiquark to form a colorless meson~\cite{Greco:2003xt,Greco:2003vf,Chen:2003qj}, that is
\begin{eqnarray}
f_M(\boldsymbol\rho,{\bf k}_\rho)=8g_M\exp\left[-\frac{\boldsymbol\rho^2}{\sigma_\rho^2}-{\bf k}_\rho^2\sigma_\rho^2\right],
\label{meson}
\end{eqnarray}
where
\begin{eqnarray}\label{meson}
\boldsymbol\rho=\frac{1}{\sqrt{2}}({\bf r}_1-{\bf r}_2),\quad{\bf k}_\rho=\sqrt{2}~\frac{m_2{\bf k}_1-m_1{\bf k}_2}{m_1+m_2},\nonumber\\
\end{eqnarray}
with $m_i$, ${\bf r}_i$ and ${\bf k}_i$ being the mass, position and momentum of quark (antiquark) $i$, respectively. The width parameter $\sigma_\rho$ in the Wigner function is related to the root-mean-square radius of the meson via
\begin{eqnarray}
\langle r_M^2 \rangle&=&\frac{3}{2}\frac{m_1^2+m_2^2}{(m_1+m_2)^2}\sigma_\rho^2\nonumber\\
&=&\frac{3}{8}\frac{m_1^2+m_2^2}{\omega  m_1m_2(m_1+m_2)},
\end{eqnarray}
where the second line follows if we use the relation $\sigma_\rho=1/\sqrt{\mu_1\omega}$ in terms of the oscillator frequency $\omega$ and the reduced mass $\mu_1=2(1/m_1+1/m_2)^{-1}$.

The probability for three quarks or antiquarks to coalescence to a baryon or an anti-baryon is similarly proportional to the quark Wigner function of the baryon, i.e.,
\begin{eqnarray}
&&f_B(\boldsymbol\rho,\boldsymbol\lambda,{\bf k}_\rho,{\bf k}_\lambda)\nonumber\\
&&=8^2g_B\exp\left[-\frac{\boldsymbol\rho^2}{\sigma_\rho^2}-\frac{\boldsymbol\lambda^2}{\sigma_\lambda^2}-{\bf k}_\rho^2\sigma_\rho^2-{\bf k}_\lambda^2\sigma_\lambda^2\right],
\label{baryon}
\end{eqnarray}
where $g_B$ is the statistical factor for three colored spin-1/2 quarks to form a colorless baryon, and
\begin{eqnarray}
{\boldsymbol\lambda}&=&\sqrt{\frac{2}{3}}\left(\frac{m_1{\bf r}_1+m_2{\bf r}_2}{m_1+m_2}-{\bf r}_3\right),\nonumber\\
{\bf k}_\lambda&=&\sqrt{\frac{3}{2}}~\frac{m_3({\bf k}_1+{\bf k}_2)-(m_1+m_2){\bf k}_3}{m_1+m_2+m_3}.
\end{eqnarray}
The width parameter $\sigma_\lambda$ is related to the oscillator frequency by $(\mu_2 \omega)^{-1/2}$, with $\mu_2=(3/2)[1/(m_1+m_2)+1/m_3]^{-1}$. The root-mean-square radius of a baryon or an antibaryon is then given by
\begin{eqnarray}
&&\langle r_B^2 \rangle\nonumber\\
&&=\frac{1}{2}\frac{m_1^2(m_2+m_3)+m_2^2(m_3+m_1)+m_3^2(m_1+m_2)}{\omega(m_1+m_2+m_3)m_1m_2m_3}.\nonumber\\
\end{eqnarray}

For the mesons $K^\pm$, the baryons $p$ and $\Lambda$, and the anti-baryons $\bar p$ and $\bar\Lambda$ considered here, their statistical factors in the quark coalescence are $g_{K^\pm}=1/36$ and $g_p=g_\Lambda=g_{\bar p}=g_{\bar\Lambda}=1/108$. The oscillator frequency $\omega$ is determined from the root-mean-square radius of the produced hadron, which is taken to be 0.6 fm for $K^\pm$, and 0.877 ${\rm fm}$~\cite{Beringer:1900zz} for both $p$ and $\Lambda$ as well as for $\bar p$ and $\bar\Lambda$. We do not consider $\pi^+$ and $\pi^-$ elliptic flows because they are affected similarly by the partonic mean fields as a result of similar elliptic flows for $u$ and $d$ quarks as well for $\bar u$ and $\bar d$ antiquarks.

\begin{figure}[h]
\centerline{
\includegraphics[width=9 cm]{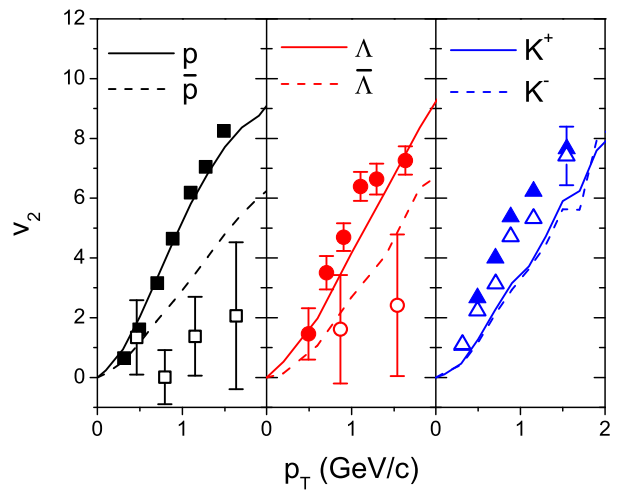}}
\caption{(Color online) Elliptic flows $v_2$ of midrapidity ($|y|<1$) $p$ and $\bar p$ (left panel), $\Lambda$ and $\bar{\Lambda}$ (middle panel), and $K^+$ and $K^-$ (right panel) at hadronization as functions of transverse momentum for $g_V=G/6$. Experimental data are taken from Refs.~\cite{Adamczyk:2013gw}.}
\label{v2-proton}
\end{figure}

\begin{figure}[h]
\centerline{
\includegraphics[width=9 cm]{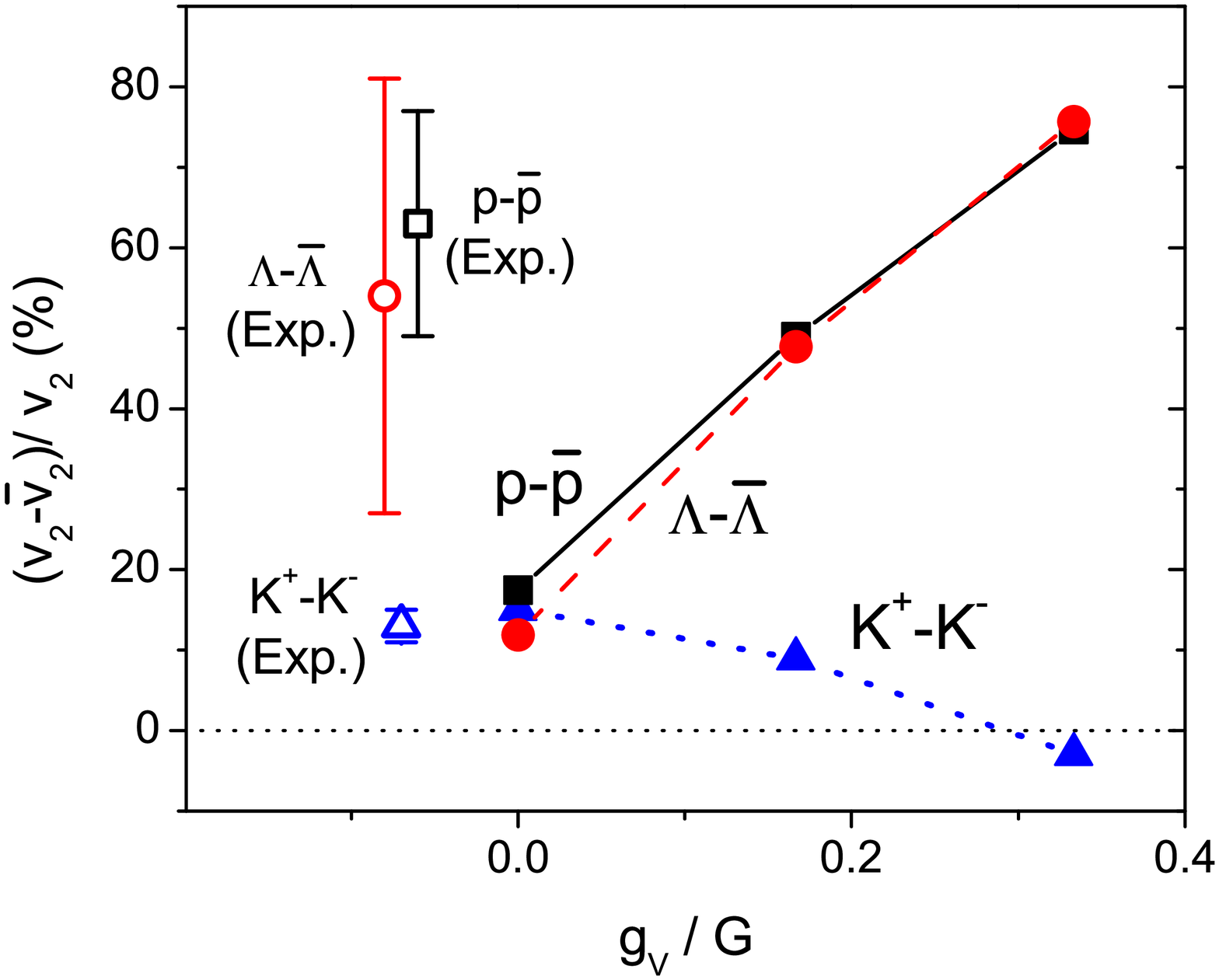}}
\caption{(Color online) Relative differences between integrated $v_2$ of mid-rapidity ($|y|<1$) $p$ and $\bar p$~(solid squares), $\Lambda$ and $\bar\Lambda$~(solid circles), and $K^+$ and $K^-$~(solid triangles) at hadronization for several values of the isoscalar vector coupling. Experimental data from Refs.~\cite{Kumar:2011us,Mohanty:2011nm} are shown by open symbols in the left side.}
\label{splittings}
\end{figure}

In Fig.~\ref{v2-proton}, we show by solid and dashed lines, respectively, the $v_2$ of $p$ and $\bar p$ (left panel), $\Lambda$ and $\bar\Lambda$ (middle panel), and $K^+$ and $K^-$ (right panel) at hadronization as functions of transverse momentum for $g_V=G/6$. It is seen that the quark coalescence leads to a larger hadron $v_2$ than the quark $v_2$ at same transverse momentum. Furthermore, the $v_2$ of $p$, $\Lambda$, and $K^+$ are respectively larger than those of $\bar p$, $\bar\Lambda$, and $K^-$, leading to the relative differences between their integrated $v_2$, $[v_2({\rm particle})-v_2({\rm antiparticle})]/v_2({\rm particle})$, of about 49, 48, and 9\%, respectively, as shown by solid symbols in Fig.~\ref{splittings} for $g_v/G=1/6$, compared with 63$\pm$14, 54$\pm$27, and 13$\pm$2\% measured in experiments shown by open symbols in the left side of Fig.~\ref{splittings}. We note that although the integrated $v_2$ of up and down antiquarks is almost zero at hadronization as shown in Fig.~\ref{v2}, the $v_2$ of $\bar p$ is slightly positive. This is due to the fact that the scaled antiproton $v_2$ as a function of the scaled transverse momentum, i.e., divided by the number of constituent antiquarks in an antiproton, is larger than that of antiquark $v_2$ as a result of appreciable eccentricity at hadronization in heavy ion collisions at $\sqrt{s}=7.7$ GeV, contrary to the case in collisions at the much higher energy of $\sqrt{s}=200$ GeV~\cite{Chen:2006vc,Molnar:2005wf}.

The dependence of the relative difference between integrated particle and antiparticle $v_2$ on the vector coupling $g_V$ is shown in Fig.~\ref{splittings}. Besides the value $g_V/G=1/6$, two other values of 0 and 0.73, the latter from the Fierz transformation, are also used. It is seen that without vector interactions, the $v_2$ of $p$, $\Lambda$ and $K^+$ are larger than those of $\bar p$, $\bar{\Lambda}$ and $K^-$, respectively, by about 15\%, since the $v_2$ of light quarks is slightly larger than that of light antiquarks as shown in Figs.~\ref{v2} and \ref{v2-pt}. With increasing strength of the vector coupling, the relative integrated $v_2$ differences between $p$ and $\bar p$ as well as between $\Lambda$ and $\bar\Lambda$ increase almost linearly. An opposite behavior is, however, seen for the relative $v_2$ difference between $K^+$ and $K^-$, i.e., it decreases with increasing strength of the vector coupling, and this is due to the fact that the vector mean field, which acts similarly on light and strange (anti-)quarks, reduces the effect due to different spatial eccentricities of quarks and antiquarks.

\section{summary}\label{summary}

We have studied the effect of partonic mean fields on the elliptic flows of quarks and antiquarks in a brayon-rich quark matter by
using a transport model based on the NJL model. For the scalar mean field, which is attractive for both quarks and antiquarks, it
leads to a similar reduction of the quark and antiquark $v_2$ as first found in Ref.~\cite{Plumari:2010ah}. The vector mean field,
on the other hand, has very different effects on quarks and antiquarks in the baryon-rich matter as it is repulsive for quarks and
attractive for antiquarks. The time component of the vector mean field turns out to have the strongest effect, resulting in a
significant splitting of the quark and antiquark $v_2$ as a result of enhanced quark $v_2$ and suppressed antiquark $v_2$. The
space component of the vector mean field has, however, an opposite effect; it suppresses $v_2$ of quarks and enhances
that of antiquark, although relatively small and appearing later in the partonic stage compared to that of the time component of
the vector mean field. Using the quark coalescence model, we have further studied the elliptic flows of $p$, $\Lambda$,
and $K^+$ and their antiparticles produced from the baryon-rich quark matter and found that the differences between particle and
antiparticle elliptic flows are appreciable as a result of the different quark and antiquark $v_2$. The magnitude of the relative
integrated $v_2$ difference between particles and their antiparticles depends on the strength of the vector coupling.
As shown in Fig.~\ref{splittings}, although using a larger vector coupling in the partonic matter can describe the
$p$ and $\bar p$ as well as the $\Lambda$ and $\bar\Lambda$ relative $v_2$ differences that were measured in
experiments by the STAR Collaboration~\cite{Adamczyk:2013gw}, it fails to reproduce the measured
relative $v_2$ between $K^+$ and $K^-$. This is not surprising since other effects that can lead to the splitting of the
elliptic flow of particles and their antiparticles have not been included in the present study. For example,
we have not included the chemical reactions of partons, such as the quark-antiquark creation and annihilation, and hadronic mean-field effects. Both are expected to also lead to a splitting of the quark and antiquark $v_2$ as shown in Ref.~\cite{Steinheimer:2012bn} based on the ideal hydro+UrQMD hybrid model that assumes local thermal and baryon chemical equilibrium in the initial stage of heavy ion collisions, and in Ref.~\cite{Xu:2012gf} based on the AMPT model using empirically determined hadronic potentials. Although a quantitative determination of the partonic vector interaction requires a more complete study that includes above mentioned effects, the present study has clearly shown that the splitting of quark and antiquark elliptic flow and thus that of particles and their antiparticles is sensitive to the strength of the partonic vector interaction. Our results therefore indicate for the first time that studying the elliptic flow in heavy ion collisions at BES energy can potentially allow for the determination of the partonic vector interaction in baryon-rich QGP and thus the equation of state of QGP at finite baryon chemical potential.

Finally, the $v_2$ of $\pi^+$ and $\pi^-$ are the same in the present study because we have not included the isovector
part of partonic mean fields. As shown in Ref.~\cite{Xu:2012gf}, including the isovector hadronic mean fields indeed leads to a
splitting of the $v_2$ of $\pi^+$ and $\pi^-$, although a factor of five smaller than the measured value. It is thus also of great interest to study the effect of quark isovector mean fields on the $\pi^+$ and $\pi^-$ $v_2$ difference and compare it with that due to the chiral magnetic effect suggested in Ref.~\cite{Burnier:2011bf}.

\section*{Acknowledgements}

We thank Volker Koch and Jan Steinheimer for helpful communications. This work was supported in part by the U.S. National Science Foundation under Grant No. PHY-1068572, the Welch Foundation under Grant No. A-1358, and the ERC-StG under the Grant QGPDyn No. 259684.

\end{document}